\documentclass[twocolumn,aps,epsfig,showpacs,floatfix]
{revtex4}
\usepackage{epsfig}
\usepackage{graphicx}
\usepackage{dcolumn}

\begin{document}

\title{Systematics of heavy-ion fusion hindrance at extreme sub-barrier
energies}
\author{C.~L.~Jiang}
\author{B.~B.~Back}
\author{H.~Esbensen}
\author{R.~V.~F.~Janssens}
\author{K.~E.~Rehm}
\affiliation{Physics Division, Argonne National Laboratory, Argonne, IL 60439}

\date{\today}

\begin{abstract}
The recent discovery of hindrance in heavy-ion induced fusion reactions at
extreme sub-barrier energies represents a challenge for theoretical models.
Previously, it has been shown that in medium-heavy systems, the onset of
fusion hindrance depends strongly on the "stiffness" of the nuclei in the
entrance channel. In this work, we explore its dependence on the total mass
and the $Q$-value of the fusing systems and find that the fusion hindrance
depends in a systematic way on the entrance channel properties over a wide
range of systems.

\end{abstract}

\maketitle

During the last thirty years of sub-barrier fusion studies, three important
observations have been made: 1) the discovery of sub-barrier fusion
enhancement associated with couplings to the intrinsic excitations of the
participating nuclei \cite{sum1a, sum1b, sum1c, sum1d}; 2) measurements of
the spin-distributions of the fused-compound nuclei and their theoretical
description \cite {sum2}; 3) the introduction of the concept of barrier
distributions and their subsequent detailed measurements \cite {sum3a,
sum3b}. In these studies it has been found that new representations of the
fusion cross sections, such as the spin distribution $d\sigma (l) /dl$, the
moments $\langle l \rangle$ and $ \langle l^2 \rangle$ of this distribution,
as well as the quantity $d^2(E \sigma) /dE^2$ associated with the
distribution of fusion barriers \cite{sum3a}, are essential for exposing
pertinent features of the data. In general, the coupled-channels theory,
when using appropriate ion-ion potentials, is able to describe the fusion
cross section for moderately heavy systems down to an energy just below the
interaction barrier. Recently, it was found that to reproduce fusion cross
sections at above barrier energies it was necessary to increase the
diffuseness parameter to values larger than those derived from elastic
scattering data \cite{newton}. This effect is possibly associated with the
opening of the deep-inelastic reaction channel. For very heavy systems, it
is well known that dynamical effects hinder the formation of a compound
nucleus leading to more complicated exit channels such as deep inelastic and
quasi-fission processes. The dynamical hindrance of fusion in such systems
has been described as a diffusion process which may eventually reach the
configuration of the compound nucleus \cite{Swiatecki}.

Recently, a new phenomenon of hindrance in heavy-ion fusion reactions has
been found in medium-heavy systems \cite {jiang0, jiang1, jiang2,jiang3}.
This hindrance occurs at extreme sub-barrier energies whereas the fusion
cross section at near barrier energies agrees fairly well with standard
coupled-channels calculations. At present, the exploration of this hindrance
phenomenon is only in its initial stage; the underlying physics reason is
still unknown. Several colliding systems have been measured down to very low
cross section levels. In addition, many existing data have been reanalyzed
in order to uncover systematic trends. Thus, it has been found that the
nuclear structure of the fusing nuclei plays a decisive role for the onset
energy for the hindrance in medium-heavy systems~\cite{jiang2,jiang3}. In
the present paper, we study the dependence of the hindrance on the mass, and
by extension also on the $Q$-value of the fusing systems over a wide range
of projectile-target combinations. In general, the $Q$-value becomes less
negative with decreasing mass, and even positive for the lightest systems.
We note that the possible occurance of fusion hindrance in the lightest
nuclei is of great astrophysical interest.

\begin{figure} 
\epsfig{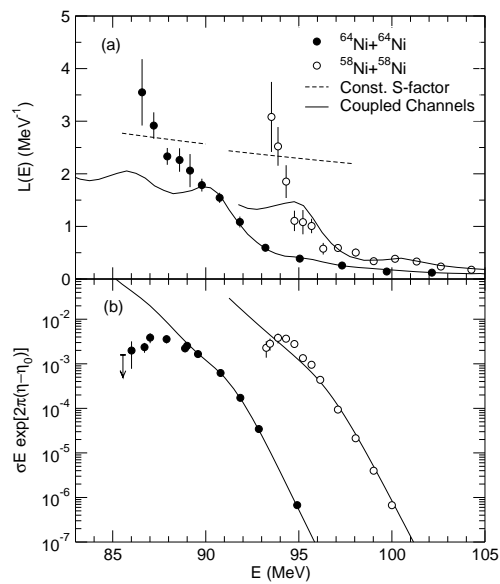}
\caption{Comparison of the logarithmic derivative and $S$-factor
  representations of the fusion cross section for the systems,
  $^{58}$Ni+$^{58}$Ni~\cite{Beck58} and $^{64}$Ni+$^{64}$Ni~\cite{jiang2}.
  The dashed curves correspond to a constant $S$-factor, whereas the solid
  curves display results of coupled-channels calculations. The $L(E)$ data
  were obtained from a fit to the cross sections at three consecutive beam
  energies.}
\label{fig1}
\end{figure}

In order to be able to recognize the hindrance in the rapidly varying
sub-barrier fusion cross sections, we have earlier studied the effect in
terms of two representations, which are not often used in heavy-ion fusion
studies. These are the logarithmic derivative, $L(E)=d\ln(\sigma E)/dE$ and
the $S$-factor, $S(E)=\sigma E \exp(2\pi \eta)$, where $\eta =
Z_1Z_2e^2/(\hbar v)$ is the Sommerfeld parameter
~\cite{jiang0,jiang1,jiang2,jiang3}. In Fig.~\ref{fig1} these quantities are
given for the $^{58}$Ni+$^{58}$Ni and $^{64}$Ni+$^{64}$Ni systems. The
maximum in the $S$-factor occurs at the energy, $E_s$, where the logarithmic
derivative $L(E)$ crosses the curve for a constant $S$-factor, which is
given by~\cite{jiang1}
\begin{equation}
L_{cs}(E)=\pi\eta/E=0.495Z_1Z_2\sqrt{\mu}/E^{3/2} \ \ {\rm (MeV^{-1})},
\end{equation}
where $\mu=A_1A_2/(A_1+A_2)$ (dashed curves in Fig.~\ref{fig1}). We note
that the logarithmic slope of the data, $L(E)$, intersects $L_{cs}(E)$ at a
substantially larger angle, and therefore the peak in the $S$-factor is
narrower, for $^{58}$Ni+$^{58}$Ni than for $^{64}$Ni+$^{64}$Ni. This is a
consequence of the "stiffness" of the former system. A dependence on the
"stiffness" of the fusing nuclei has been seen in many cases such as
$^{90}$Zr+$^{90,92}$Zr, $^{89}$Y (see Table I column 5) and as discussed in
Ref.~\cite{jiang0,jiang1}.

A negative fusion $Q$-value requires that there be a maximum of $S(E)$
\cite{jiang1}. This is a consequence of the fact that the cross section must
vanish at a finite center-of-mass energy corresponding to the ground state
of the fused system, {\it i.e.}, at $E=-Q$. In this limit, $L(E)=\sigma^{-1}
d\sigma/ dE + 1/E \rightarrow +\infty$, whereas $L_{cs}(E)$ remains finite.
This means that for such systems there is always an energy for which the
$S$-factor has a maximum. This occurs when $L(E)$ becomes equal to
$\pi\eta/E$, a condition that is always fulfilled, since $\pi\eta/E$ is
finite near and above $E=-Q$, whereas $L(E)=d\ln(\sigma E)/dE$ approaches
infinity as $E\rightarrow-Q$.

For positive $Q$-value systems, however, a maximum may not develop, because
both $L(E)$ and $L_{cs}(E)$ become infinite in the limit of $E=0$. If $L(E)$
does not grow faster than $L_{cs}(E)$ with decreasing energy, it may not
cross $L_{cs}(E)$ for any positive value of $E$. It is, therefore, of
interest to study the systematics of the sub-barrier fusion hindrance over a
wide range of systems, including some with positive $Q$-value, as is the
case mainly in fusion between lighter nuclei.

\begin{figure} 
\epsfig{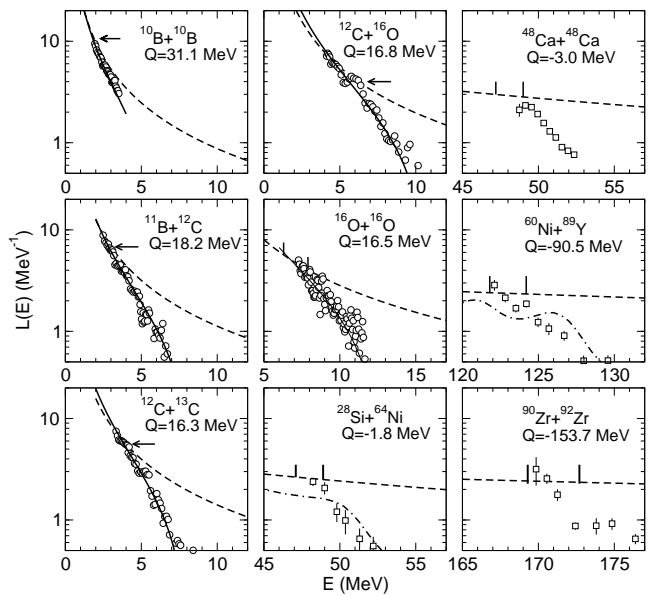}
\caption{Logarithmic derivative representations for a range of systems.
  The dashed curves correspond to a constant $S$-factor, whereas the
  dashed-dotted curves display results of coupled-channels calculations, and
  the solid curves represent a fit to the data using the function
  $a+b/E^{3/2}$. The range of $E_s$ values are indicated by vertical line
  segments for heavy systems. For the four lightest systems only lower
  limits of $L_s$ can be derived (shown as arrows). The data are taken from
  Refs.: $^{10}$B+$^{10}$B \cite{b10}, $^{11}$B+$^{12}$C ,
  $^{12}$C+$^{13}$C, $^{12}$C+$^{16}$O, $^{16}$O+$^{16}$O \cite{stok},
  $^{48}$Ca+$^{48}$Ca \cite{trot}, $^{60}$Ni+$^{89}$Y \cite{jiang0}, and
  $^{90}$Zr+$^{92}$Zr \cite{Keller}. }
\label{fig2}
\end{figure}

The expected dependence on the $Q$-value of the system appears to be borne
out by data. The systematics of the logarithmic derivative $L(E)$ of fusion
excitation functions is illustrated in Fig.~\ref{fig2} for a number of
systems ranging from $^{10}$B+$^{10}$B to $^{90}$Zr+$^{92}$Zr.  The
logarithmic derivatives are represented by open circles for five-point
derivatives, whereas the open squares were obtained by a fit to three
consecutive data points. We observe that $L(E)$ for all systems increases
with decreasing energy. The dashed curves represent the logarithmic slopes
corresponding to a constant $S$-factor (Eq. 1). In an earlier study of
fusion between "stiff" nuclei~\cite{jiang1}, which did not include systems
lighter than $^{16}$O+$^{144}$Sm, we found that the $S$-factor maximum
systematically occurred at a value of $L_s$=2.33 MeV$^{-1}$ corresponding to
\begin{equation}
E_s^{\it ref} = 0.356\left(Z_1Z_2\sqrt{\mu}\right)^{\frac{2}{3}} \ \
\rm{(MeV)}.
\label{eqn2}
\end{equation}
Studying the full range of systems, we observe that the crossing point,
$E_s$, for lighter systems, which have increasingly positive $Q$-values,
indeed occurs at larger values of $L(E)$. For the lightest systems, the
logarithmic derivatives of the data intersect the constant $S$-factor curve
at a small angle and it is, therefore, difficult to accurately estimate
$E_s$. Consequently, we have used fits to the data with the expression
$a+b/E^{3/2}$ (solid curves), $a$ and $b$ being adjustable parameters, to
obtain a less subjective estimate of $E_s$. The results are given in Table
I. Relatively large error bars are, however, assigned to the resulting $E_s$
values and, for the lightest systems, only upper limits are given, because
of the inaccuracy of this procedure. We also observe that the value of the
logarithmic slope, $L(E)$, obtained by coupled-channels calculations for
heavy systems (dashed-dotted curves in Fig. 2) saturates at a value of
$\sim$ 1.5 - 2.0 MeV$^{-1}$, much lower than measured. It has been shown
that coupled-channels calculations using reasonable ion-ion potentials are
unable to reproduce the extreme sub-barrier behavior~\cite{jiang1}.

\begin{table*}[bt] 
\caption{This table lists the parameter $Z_1Z_2\sqrt{\mu}$,
  the energy $E_s$ and the logarithmic derivative, $L_s=L(E_s)$, that
  characterize the maximum of the {\it S}-factor for different systems. Also
  given are the $(dL/dE)_{exp}$ and $(dL/dE)_{cs}$, corresponding to the
  measured and the constant $S$-factor curves at $E_s$. $R$ is the ratio
  $(dL/dE)_{exp}/(dL/dE)_{cs}$, $Q$ is the fusion $Q$-value, and $V_{Bass}$
  is the height of the Bass barrier~\cite{Bass}. Systems in categories I and
  II exhibit a clear maximum in the $S(E)$ curve for "stiff" and "soft"
  systems, respectively. A maximum has not quite been reached for systems in
  category III and IV. Extrapolated values of $E_s$ and $L_s$ etc. are
  listed for category III, and only upper limits for $E_s$ (lower limits of
  $L_s$) are included for most category IV systems.  Uncertainties are given
  in parentheses for $E_s$, $(dL/dE)_{exp}$ and $R$. In cases where only
  upper limits for $E_s$ can be given the values of $(dL/dE)_{\it exp},
  (dL/dE)_{\it cs}$, and $R$ correspond to the crossing points obtained from
  the fit to the data. Uncertainties for $L_s$ and $(dL/dE)_{cs}$ can be
  obtained from the uncertainties on $E_s$ with the constant $S$-factor
  formula.}

\begin{ruledtabular}
\begin{tabular} {cccccccccc}
System &
 $Z_1Z_2\sqrt{\mu}$ & $E_s$ & $L_s$ & $(dL/dE)_{exp}$ & $(dL/dE)_{cs}$ & $R$ 
 & $Q$ & $V_{Bass}$& Ref. \\
 &       & (MeV) & (MeV$^{-1}$) & (MeV$^{-2}$) &  (MeV$^{-2}$) &    & (MeV) 
 & (MeV) &  \\
\tableline
Category I\\
\tableline
$^{90}$Zr+$^{90}$Zr  & 10733 &  175(1.8)  &  2.29 & -1.61(0.16)  & -0.020 & 
81.9(8.2) & -157.35  & 195.3 & \cite{Keller} \\
$^{90}$Zr+$^{89}$Y   & 10436 &  171(1.7)  &  2.31 & -1.12(0.08)  & -0.020 & 
55.1(4.4) & -151.53  & 190.1 & \cite{Keller} \\
$^{90}$Zr+$^{92}$Zr  & 10792 &  171(1.7)  &  2.40 & -0.84(0.07)  & -0.021 & 
39.0(3.6) & -153.71  & 184.4 & \cite{Keller} \\
$^{58}$Ni+$^{58}$Ni  &  4222 &   94(0.9)  &  2.29 & -1.64(0.31)  & -0.036 & 
44.9(8.6) & -66.122  & 102.0 & \cite{Beck58} \\
$^{60}$Ni+$^{89}$Y   &  6537 &  123(1.2)  &  2.38 & -0.80(0.19)  & -0.029 & 
27.5(6.6) & -90.497  & 136.5 & \cite{jiang0} \\
$^{32}$S+$^{89}$Y    &  3026 &  72.6(0.7) &  2.42 & -0.58(0.15)  & -0.050 & 
11.5(3.0) & -36.597  & 79.8  & \cite{mukh}   \\
\tableline
Category II\\
\tableline
$^{64}$Ni+$^{100}$Mo & 7343  &  121(1.2)  &  2.74 & -0.57(0.09)  & -0.034 & 
17.0(2.7) & -92.287  & 143.3 & \cite{jiang3}\\ 
$^{64}$Ni+$^{64}$Ni  & 4435  &  87.3(0.9) &  2.69 & -0.35(0.02)  & -0.046 & 
 7.7(0.5) & -48.783  & 98.1  & \cite{jiang2} \\
\tableline
Category III\\
\tableline
$^{48}$Ca+$^{48}$Ca & 1960 & 48.1(0.9) & 2.90 & -0.59(0.03) & -0.090 & 
6.5(0.5) & -2.988   & 50.1 & \cite {trot}\\ 
$^{28}$Si+$^{64}$Ni & 1729 & 47.3(0.9) & 2.57 & -0.70(0.12) & -0.080 & 
8.7(1.7) & -1.783   & 50.8 & \cite {stef1}\\ 
$^{16}$O+$^{76}$Ge & 930.5 & 27.6(0.8) & 3.17 & -0.36(0.05)  & -0.172 &
2.1(0.2) & 10.506   & 32.5 & \cite {agui}\\
\tableline
Category IV\\
\tableline
$^{16}$O+$^{16}$O & 181.0 & 7.1(0.8)     & 4.7(0.7) & -1.7(0.2)  & -1.0  & 
1.7(0.2) & 16.542  & 8.2  & \cite {wu,thom,stok,spin,hulke}\\
$^{12}$C+$^{16}$O & 125.7 & $<$6.2     & $>$4.0  & -3.0  & -2.2  & 
1.4 & 16.756  & 6.0  & \cite {stok,cuje,swit,c12}\\
$^{12}$C+$^{14}$N & 106.8 & $<$5.0     & $>$4.7  & -4.1  & -3.2  & 
1.3 & 15.074  & 5.2  & \cite {stok}\\
$^{12}$C+$^{13}$C &  89.9 & $<$4.0     & $>$5.6  & -3.9  & -2.7  & 
1.4 & 16.318  & 4.3  & \cite {stok}\\ 
$^{11}$B+$^{12}$C & 71.9  & $<$3.0     & $>$6.8  & -8.8  & -7.6  & 
1.2 & 18.198  & 3.5  & \cite {stok}\\ 
$^{10}$B+$^{10}$B & 55.9  & $<$1.9     & $>$10.6 & -26.6 & -24.4 & 
1.1 & 31.144  & 2.9  &\cite {b10}\\ 
\end{tabular}
\end{ruledtabular}
\end{table*}

\begin{figure}[hbt] 
\epsfig{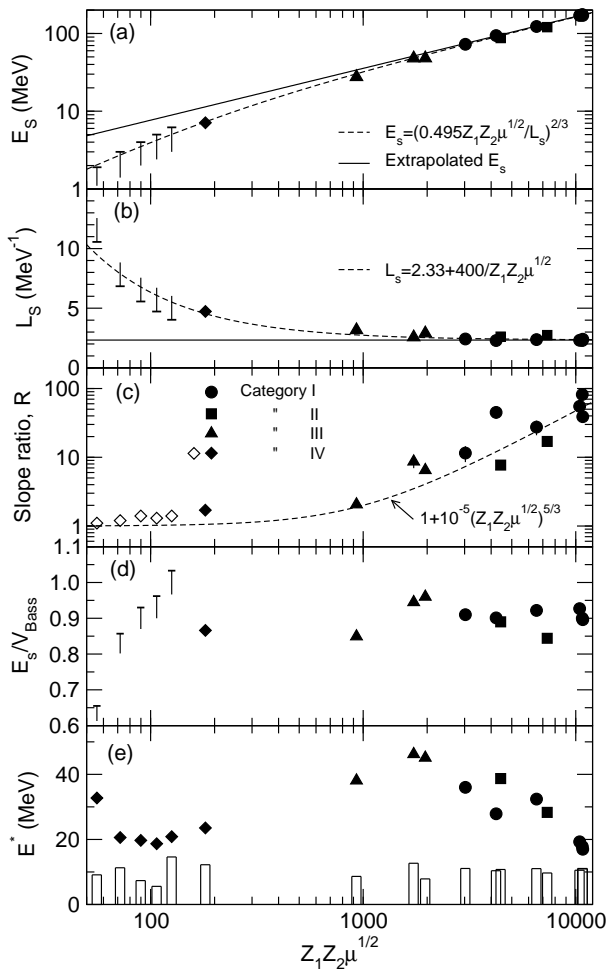}
\caption{(a) Experimental values (symbols) and limits (horizontal bars) are
  shown as a function of $Z_1Z_2\sqrt{\mu}$ for $E_s$ (panel a), $L_s$
  (panel b), logarithmic slope ratio, $R =(dL/dE)_{exp}/(dL/dE)_{cs}$ at
  $E_s$ (panel c), ratio $E_s/V_{Bass}$ (panel d), and $E^*=E_s+Q$ (panel
  e), which also shows effective particle emission thresholds (histogram).
  Solid lines in panels a) and b) correspond to $L_s$=2.33 MeV$^{-1}$,
  whereas dashed curves represent empirical trends of the data. Open
  diamonds in panel c) represents the slope ratio at the crossing point
  obtained from the fit to the data.
}
\label{fig3}
\end{figure}
The systematics of sub-barrier hindrance is illustrated in Fig. 3. Here, the
derived values of $E_s$ and $L_s=L(E_s)$ are plotted as a function of the
parameter $Z_1Z_2\sqrt{\mu}$ in panels a) and b), respectively. Aside from
local deviations of $L_s$ from the value of 2.33 MeV$^{-1}$ in medium-heavy
systems (of the order of $\sim 10\%$, arising from nuclear structure
effects) $L_s$ clearly starts deviating from this value in lighter systems.
The corresponding $E_s$ values also fall below the $E_s^{\it ref}$
systematics (solid curve) given in Eq. 2. A purely empirical expression
\begin{equation}
L_s^{\it emp}=2.33 + 400/(Z_1Z_2\sqrt{\mu}) \ \ {\rm (MeV^{-1})}
\label{eqn3}
\end{equation}
(dashed curve in Fig. 3b) is seen to provide a good approximation to the
experimental data, and it reproduces the asymptotic value of 2.33 MeV$^{-1}$
observed earlier for heavy systems with $Z_1Z_2\sqrt{\mu}>2500$. The
corresponding curve for $E_s^{emp}$ obtained from Eqs. 1 and 3, namely
\begin{equation}
E_s^{\it emp}=(0.495Z_1Z_2\sqrt{\mu}/L_s^{\it emp})^{2/3} \ \ ({\rm MeV}),
\label{eqn4}
\end{equation}
is seen also to reproduce the experimental values in Fig. 3a. These two
equations thus represent the overall systematics for the onset of
sub-barrier fusion hindrance. This systematics appears to be correlated with
the parameter $Z_1Z_2\sqrt{\mu}$ in the simple fashion expressed in Eqs. 3
and 4, but it should be kept in mind that both the $Q$-value and the fusion
(interaction) barrier vary smoothly, although not quite monotonically, with
this parameter. Hence, it is not possible to ascertain whether the observed
physical effect of fusion hindrance is associated with either, or with both
of these quantities.

Figure 3c presents the ratio of the logarithmic slopes, $R=(dL/dE)_{\it
exp}/(dL/dE_{\it cs})$, for the data relative to the constant $S$-factor
curve. For $Z_1Z_2\sqrt{\mu}>$ 2000, this ratio is substantially larger than
unity which means that there is a sharp intersection point between the two
curves and, consequently, a well defined, narrow maximum in the $S$-factor
curve. For $Z_1Z_2\sqrt{\mu}$ values below about 2000, the slope ratio
approaches unity which results in a less well defined intercept point. For
the lightest systems, it appears that the logarithmic slope of the data
approaches the value for a constant $S$-factor and the sub-barrier hindrance
may well disappear.

It should be emphasized that $L_{cs}(E)$ equals the logarithmic derivative
in a point charge, pure Coulomb penetration model, as long as $\eta$ is
greater than $\sim$10, which is always the case in the energy range studied
here. A relative slope, $R \gg 1$, therefore, implies that the fusion cross
section drops more rapidly than predicted in a point charge pure Coulomb
interaction model, whereas a relative slope near unity indicates that the
fusion cross section decreases at the predicted rate.

For orientation, it may also be of interest to relate the observed values of
$E_s$ to the fusion (or interaction) barrier and the $Q$-value of the fusion
process, as given in Fig. 3d and 3e as a function of the parameter
$Z_1Z_2\sqrt{\mu}$. Since the cross section must vanish at an energy of
$E=-Q$ (for $Q<0$) or $E=0$ (for $Q>0$), the larger of these two values
represents a lower bound for $E_s$. On the other hand, one may consider the
fusion barrier, taken here from the Bass prescription~\cite{Bass},
$V_{Bass}$, as an upper bound for $E_s$. In Fig. 3d, the experimental values
of $E_s$ do not appear to have a simple or fixed relation to $V_{\it Bass}$:
the onset of sub-barrier hindrance, $E_s$, occurs at an energy between 5\%
and 35\% below the fusion barrier. Furthermore, relatively large
fluctuations between systems with similar values of the parameter
$Z_1Z_2\sqrt{\mu}$ are present, some of which are clearly related to the
structure of the fusing nuclei~\cite{jiang2,jiang3}.

A potential cause of fusion hindrance at sub-barrier energies could be the
rarefication of final states accessible in the fused system, which may be
expressed as the ratio of total width to the spacing of the states, {\it
i.e.} $\Gamma^{tot}/D$, in the appropriate energy regime of the compound
nucleus. In other words, a fusion reaction can only proceed if a quantum
state with the appropriate energy, spin and parity is available in the
compound nucleus. The $\Gamma^{tot}/D$ value is expected to increase
exponentially with excitation energy and will approach unity slightly above
the particle (n, p, or $\alpha$) emission threshold (binding energy +
Coulomb barrier). In order to explore this possibility, we have plotted in
Fig. 3e, the excitation energy corresponding to $E_s$, {\it i.e.}, $E_s+Q$
(solid symbols) and compared it to the particle thresholds for the systems
listed in Table I. We note that the experimental values of $E_s$ correspond
to excitation energies exceeding the particle thresholds by a wide margin in
all cases, and there does not appear to be a significant correlation between
these two quantities. Even accounting for the fact that some of the
excitation energy is bound in rotational energy does not alter this
conclusion. Although the simple explanation of rarefication of the final
states in the fusion process is appealing, it does not appear to account for
the observed hindrance phenomenon. It seems that the behavior seen in Fig.
3a-c provides indications that the hindrance phenomenon is closely related
to the entrance channel.

In conclusion, the systematics of sub-barrier fusion hindrance has been
studied over a wide range of systems from $^{10}$B+$^{10}$B to
$^{90}$Zr+$^{90}$Zr. Hindrance appears to be a general phenomenon, at least
for systems with $Z_1Z_2\sqrt{\mu} \agt 3000$. For the lightest systems
($Z_1Z_2\sqrt{\mu} \alt 200$), the logarithmic slopes of the cross section
in the sub-barrier region merge smoothly into those expected on the basis of
a constant $S$-factor ({\it i.e.} a point charge in a pure Coulomb
interaction model). Simple empirical formulae are given for both the energy
and the logarithmic slope of the cross section at which the onset of fusion
hindrance occurs. These point to an entrance channel effect as the source of
this phenomenon. Until now, all direct observations of fusion hindrance have
been made in systems with $Z_1Z_2\sqrt{\mu} \agt 3000$. It would be
interesting to study sub-barrier fusion in more systems in the range $200
\alt Z_1Z_2\sqrt{\mu} \alt 3000$, where the fusion $Q$-values change from
positive to negative values. If the fusion hindrance does indeed occur in
light systems, such as $^{12}$C+$^{12}$C, $^{12}$C+$^{16}$O and
$^{16}$O+$^{16}$O, it will strongly affect the predicted rates of
astrophysical processes, which are presently obtained by simple empirical
extrapolations from experimental data.  As yet, no satisfactory theoretical
explanation for this phenomenon has been put forth. Simple considerations in
terms of relations to the fusion barrier height or the rarefication of
compound states in the fusion channel do not appear to clarify the
situation.

This work was supported by the U.S.Department of Energy, Office of Nuclear
Physics, under contract No. W-31-109-ENG-38.


\begin{references}
\bibitem{sum1a} M. Beckerman, Physics Report, {\bf 129}, 145 (1985).
\bibitem{sum1b} M. Beckerman, Rep. Prog. Phys. {\bf 51}, 1047 (1988).
\bibitem{sum1c} K. Hagino, N. Takigawa, M. Dasgupta, D. J. Hinde and 
  J.R. Leigh, Phys. Rev. C {\bf 55}, 276 (1997).
\bibitem{sum1d} A. B. Balantekin and N. Takigawa, Rev. Mod. Phys. {\bf 70}, 77 
  (1998).
\bibitem{sum2} R. Vandenbosch, Annu. Rev. Nuc. Part. Sci. {\bf 42}, 447 (1992).
\bibitem{sum3a} N. Rowley, G. R. Satchler and P.H.Stelson, Phys. Lett.
  {\bf B254}, 25 (1991). 
\bibitem{sum3b} M. Dasgupta, D.J. Hinde, N. Rowley and A.M. Stefanini,
  Annu. Rev. Nucl. Part. Sci. {\bf 48}, 401 (1998).
\bibitem{newton} J. O. Newton {\it et al.}, Phys. Lett. {\bf B 586}, 219 
  (2004); Phys. Rev. C {\bf 70}, 024605 (2004).
\bibitem{Swiatecki}  W. J. Swiatecki, A. Trzcinska, and J. Jastrzebski, 
  Phys. Rev. C {\bf 71}, 047301 (2005).
\bibitem{jiang0} C.L. Jiang {\it et al.,} Phys. Rev. Lett. {\bf 89}, 052701
  (2002).
\bibitem{jiang1} C.L. Jiang, H.Esbensen, B.B. Back, R.V.F. Janssens and
  K.E. Rehm, Phys. Rev. C. {\bf 69}, 014604 (2004).
\bibitem{jiang2} C.L. Jiang {\it et al.,} Phys. Rev. Lett. {\bf 93},
  012701 (2004).
\bibitem{jiang3} C.L. Jiang {\it et al.,} Phys. Rev. C {\bf 71}, 044613 (2005).
\bibitem{Beck58} M. Beckerman {\it et al}., Phys. Rev. C {\bf 23}, 1581 (1982).
\bibitem{b10} M.D. High and B. Cujec, Nucl. Phys. {\bf A259}, 513 (1976).
\bibitem{stok} R.G. Stokstad {\it et al.,} Phys. Rev. Lett. {\bf 37} 888
 (1976).
\bibitem{trot} M. Trotta {\it et al.,} Phys. Rev. C {\bf 65}, 011601 (2002).
\bibitem{Keller} J. G. Keller {\it et al}., Nucl. Phys. {\bf A452}, 173 (1986).
                 Phys. Lett. B {\bf 254}, 25 (1991).
\bibitem{Bass} R.Bass, Nucl. Phys. {\bf A231}, 45 (1974).
\bibitem{mukh}  A. Mukherjee, {\it et al}., Phys. Rev. C {\bf 66}, 034607
 (2002).
\bibitem{stef1} A.M. Stefanini {\it et al.,} Nucl. Phys. {\bf A456}, 509
 (1986).
\bibitem{agui} E.F. Aguilera, J.J. Kolata and R.J. Tighe, Phys. Rev. C {\bf 52}, 3103 (1995).
\bibitem{spin} H. Spinka {\it et al.,} Nucl. Phys. {\bf A233} 456 (1974).
\bibitem{wu} S.C. Wu and C.A. Barnes, Nucl. Phys. {\bf A422}, 373 (1984).
\bibitem{thom} J. Thomas {\it et al.,} Phys. Rev. C
 {\bf 31}, 1980 (1985).
\bibitem{hulke} G. Hulke {\it et al.}, Z. Phys. {\bf A297}, 161 (1980).
\bibitem{cuje} B. Cujec {\it et al.,} Nucl. Phys. {\bf A266} 461 (1976).
\bibitem{swit} Z.E. Switkowski {\it et al.,} Nucl. Phys. {\bf A274} 202
 (1976).
\bibitem{c12} J.R. Patterson, H. Winkler and C. S. Zaidins, Astrophys. J. 
  {\bf 157}, 367 (1969).
\end{references}
\end{document}